\documentclass[10pt,aps,prd,showpacs,superscriptaddress,twocolumn]{revtex4}
\usepackage{amsfonts}
\usepackage{amssymb}
\usepackage{amsmath}
\usepackage{MnSymbol}
\usepackage{graphicx}

\date{\today}

\begin{document}

\title{Reply to ``Comment on `Quantum versus classical instability of scalar fields in curved backgrounds'"}

\author{George E.\ A.\ Matsas}
\affiliation{Instituto de F\'\i sica Te\'orica, Unesp, R.\ Dr.\ Bento Teobaldo Ferraz 271, 01140-070, 
S\~ao Paulo, SP, Brazil}

\author{Raissa F.\ P.\ Mendes}
\affiliation{Instituto de F\'\i sica Te\'orica, Unesp, R.\ Dr.\ Bento Teobaldo Ferraz 271, 01140-070, 
S\~ao Paulo, SP, Brazil}

\author{Daniel A.\ T.\ Vanzella}
\affiliation{Instituto de F\'\i sica de S\~ao Carlos,
USP, C.\ P.\ 369, 13560-970, S\~ao Carlos, SP, Brazil}

\begin{abstract}
{\bf We show that the reasoning which led the author of Ref.~\cite{K} 
(arXiv:1310.6252) to reach his conclusions relies on an incorrect criterion 
for the existence of normalizable bound solutions.} We reinforce that the 
general result derived in the Appendix of our paper~\cite{MMV} (arXiv:1310.2185), 
namely, that ``{\bf there are no tachyonic (i.e., unstable) modes for 
minimally coupled scalar fields in asymptotically flat spherically symmetric 
static spacetimes containing no horizons}" is indeed correct.
\end{abstract}
\pacs{{04.62.+v}}

\maketitle

Let us begin pointing out the mistake which led the author of Ref.~\cite{K}
to reach the incorrect conclusion that there could exist tachyonic (i.e., 
unstable) modes for minimally coupled scalar fields in asymptotically 
flat spherically symmetric static spacetimes containing no horizons.
The whole point concerns whether or not the following equation 
\begin{equation}
-\psi'' + V_{\rm eff} \,\psi = - \Omega^2 \psi,\;\;\;\;\;\;    V_{\rm eff} \equiv {r''}/{r},
\label{key}
\end{equation}  
satisfying 
$$
(i)\; r'(x) >0, \;\;\; x \in [0, +\infty),\;\;\; 
{\rm and} \;\;\;
(ii)\; \lim_{x\to+\infty} r'(x) = 1,
$$
(where $``\, ' \," \equiv d/dx$)
admits normalizable bound solutions for some  real number 
$\Omega$. 
According to the author 
of Ref.~\cite{K} there should exist some (deep enough) 
$V_{\rm eff}$ satisfying the conditions above which would 
admit normalizable bound solutions. In order to ``prove" this, he writes
down a particular spherically symmetric $V_{\rm eff}$ and states that it would admit such
solutions simply based on the fact that $V_{\rm eff}$ satisfies  
\begin{equation}
\int_{x_1}^{x_2} dx \sqrt{-V_{\rm eff}} \geq \pi/2,
\label{L}
\end{equation}
where $x_1$ and $x_2$ are the ``classical" turning points.
{\bf It happens, however, that inequality~(\ref{L}) is not
a sufficient condition for the existence of normalizable bound 
solutions.} 
This can be attested even by using well-known spherically symmetric 
potentials as the Yukawa one: $-V_0 a e^{-x/a}/x$ (with $V_0,a>0$). 
This potential does satisfy inequality~(\ref{L}) for
$\pi/8\leq V_0 a^2\leq 1.6798$ but has no bound solution for parameters in this range
(see, e.g., Refs.~\cite{C,PE}). {\bf In fact, beyond the author's claim
based on the na\"ive use of Eq.~(\ref{L}), he does not exhibit any bound solution 
for the potential he writes down. Indeed, it would be impossible.}

A precise proof that Eq.~(\ref{key}) does not admit normalizable bound 
solutions was shown in the Appendix of Ref.~\cite{MMV}. Notwithstanding,
in order to dismiss any remaining doubts, we offer in the appendix below
an even more detailed step-by-step proof.  As a consistency check, we recall 
that the quantum and classical instability regions exhibited in 
Refs.~\cite{LMV} and~\cite{panietal} for spherical stars coincide with each 
other and are in precise agreement with the nonexistence of 
tachyonic modes for minimally coupled scalar fields. It is also worthwhile 
to mention that contrary to the claims of Ref.~\cite{K}, the fact that there 
is no tachyonic instability for minimally coupled scalar fields in asymptotically 
flat spherically symmetric static spacetimes containing no horizons does 
not imply anything to the Jeans instability, since in the latter case 
the effective potential is less constrained than in the former one.

\appendix*
\section{}

We start by recalling the explicitly-derived  result of the Appendix of Ref.~\cite{MMV} that the solution of 
the equation
$$
-\psi_0''(x)+V(x) \psi_0(x) = 0
$$
satisfying $\psi_0(0)=0$ and $\psi_0'(0)>0$ is monotonically increasing 
for $x\geq 0$ [where $V(x)$ is given in Eq.~(A2) of the Appendix of Ref.~\cite{MMV} and can 
be cast as $V(x)=r''(x)/r(x)$ -- hence, 
$r(x)$ itself plays the role of the desired solution, $\psi_0(x)= C r(x)$, $C>0$ a constant]. Now, we prove 
the following theorem: 
\begin{description} 
\item $\bullet$ {\bf  Theorem:} If the solution~\footnote{Here we only assume that solutions
are absolutely continuous functions with piecewise continuous first derivative. This allows  terms  containing
Dirac's ``$\delta$-function'' to appear in  the  potential $V$.} of the equation
\begin{equation}
-\psi_0''(x)+V(x) \psi_0(x) = 0
\label{Eqpsi0}
\end{equation}
satisfying $\psi_0(0)=0$ and $\psi_0'(0)\geq 0$ is monotonically increasing 
for $x> 0$, then the equation
\begin{equation}
\hskip 1.0cm
-\psi_\Omega''(x)+V(x) \psi_\Omega(x) = -\Omega^2 \psi_\Omega(x),\; \Omega\in \mathbb R^\ast,
\label{EqpsiOmega}
\end{equation}
has {\it no} non-vanishing solution satisfying both $\psi_\Omega(0)=0$ and $\lim_{x\to \infty}\psi_\Omega(x)=0 $ 
[i.e., the potential $V$ admits {\it no} normalizable
``bound'' solutions; recall that the condition $\psi_\Omega(0)=0$ comes from the fact that the radial part
of the field is $\psi_\Omega(x)/r(x)$ and $r(x\to 0)\to 0$].
\end{description}
{\it Proof:} Let us prove this result by {\it reductio ad absurdum}, by assuming there is a non-vanishing 
solution of the latter equation
with the desired properties. Then, 
multiply Eq.~(\ref{Eqpsi0}) by $\psi_\Omega(x)$ and Eq.~(\ref{EqpsiOmega}) by $\psi_0(x)$. Taking the
difference of the resulting equations we have:
\begin{eqnarray}
& & -\psi_0''(x) \psi_\Omega(x)+\psi_\Omega''(x) \psi_0(x) = \Omega^2 \psi_\Omega(x) \psi_0(x) \nonumber \\
&\Leftrightarrow & \frac{d}{dx}\left[\psi_\Omega'(x) \psi_0(x)-\psi_0'(x) \psi_\Omega(x)
\right]= \Omega^2 \psi_\Omega(x) \psi_0(x) \nonumber \\
&\Leftrightarrow & \psi_\Omega'(x) \psi_0(x)-\psi_0'(x) \psi_\Omega(x)
=\Omega^2 \int_0^x dy\,\psi_\Omega(y) \psi_0(y),  \nonumber \\
\label{1}
\end{eqnarray}
where we have already used $\psi_0(0)=\psi_\Omega(0)=0$. The assumption that
$\lim_{x\to \infty}\psi_\Omega(x)=0 $, together with $\psi_\Omega(0)=0$, leads to
\begin{eqnarray}
0=\lim_{x \to \infty}\psi_\Omega(x) - \psi_\Omega (0)=\int_0^\infty dx\, \psi_\Omega'(x), \nonumber
\end{eqnarray} 
which ensures that either $\psi_\Omega'$ is null (which would lead to the trivial $\psi_\Omega \equiv 0$ solution) 
or it changes sign.
Let $x^\ast $ be the {\it minimum} value of $x$  at which   $\psi_\Omega'$ changes sign. 
Note that $\psi_\Omega(x^\ast)\neq 0$ (otherwise, by the same argument above there would exist
$x^{\ast \ast}<x^{\ast}$ where $\psi_\Omega'$ would change sign, violating the
condition that $x^\ast$ is the minimum of such values). 
Therefore,
due to the linearity of Eq.~(\ref{EqpsiOmega}),
we can,   without any loss of generality, scale $\psi_\Omega$ so that $\psi_\Omega(x^\ast)= \psi_0(x^\ast)\,(>0)$.
Moreover,
even though $\psi_0'$ and $\psi_\Omega'$ may be only piecewise continuous, Eq.~(\ref{1}) shows that the combination 
in its left-hand side equals  (almost everywhere) a continuous function. All these facts together with Eq.~(\ref{1}) imply
[using the notation $\psi'(x^\ast_{\pm}) := \lim_{\epsilon \to 0}\psi'(x^\ast\pm\epsilon)$ with $\epsilon>0$]:
\begin{eqnarray}
& & \psi_\Omega'(x^\ast_+)-\psi_0'(x^\ast_+) = \psi_\Omega'(x^\ast_-)-\psi_0'(x^\ast_-) \nonumber \\
& &=\frac{\Omega^2}{\psi_0(x^\ast)} \int_0^{x^\ast} dy\,\psi_\Omega(y) \psi_0(y) > 0,
\label{2}
\end{eqnarray}
where the inequality follows from the fact that the
sign of $\psi_\Omega$ is constant in 
the interval
$(0,x^\ast]$ (for if this were not true 
then, again, there would exist $x^{\ast \ast}< x^\ast$ where $\psi_\Omega'$ changes sign). 

Recalling that $\psi_0$ is a monotonically increasing function, inequality (\ref{2}) shows 
that $\psi_\Omega'(x^\ast_+)>\psi_0'(x^\ast_+)>0$ and $\psi_\Omega'(x^\ast_-)>\psi_0'(x^\ast_-)>0$, contradicting the fact that
$\psi_\Omega'$ changes sign at $x=x^\ast$. This concludes our proof. $\square$
\\

Note that the theorem makes no {\it direct} assumption on the potential $V$; 
we only assume the existence of $\psi_0$ satisfying the hypotheses of the 
theorem. However, if additional properties are required for $V$ then we can 
strengthen the result:
\begin{description}
\item $\bullet$ {\bf  Corollary:} If $V$ is such that $\lim_{x\to \infty} V(x)=0$ and the solution of the equation
$$
-\psi_0''(x)+V(x) \psi_0(x) = 0
$$
satisfying $\psi_0(0)=0$ and $\psi_0'(0)\geq 0$ is monotonically increasing 
for $x> 0$, then the equation
$$
-\psi_\Omega''(x)+V(x) \psi_\Omega(x) = -\Omega^2 \psi_\Omega(x),\;\; \Omega\in \mathbb R^\ast,
$$
has {\it no} non-vanishing solution satisfying both $\psi_\Omega(0)=0$ and $\lim_{x\to \infty}|\psi_\Omega(x)|<\infty $.
\end{description}
{\it Proof:} This follows directly from 
the fact that the asymptotic behavior of the solution $\psi_\Omega(x)$ for large $x$ is given by
$e^{\pm \Omega x}$ in conjunction with the theorem above which shows that the behavior $e^{-|\Omega| x}$ is not acceptable. $\square$
\\

Note that these results hold true also if $\Omega^2$ is substituted by any strictly positive function. Moreover, 
the corollary above also holds for $V$ satisfying $\lim_{x\to \infty}V(x) = m^2 \geq 0$.

{\bf Acknowledgments}:
R.\ M.\ was supported by FAPESP (grant 2011/06429-3). 
G.\ M.\ and D.\ V.\ acknowledge partial support from CNPq 
and FAPESP (grant 2013/12165-4), respectively.

\end{document}